\documentclass[twocolumn,aps,prd,epsf]{revtex4}
\usepackage{epsfig}
\usepackage{dcolumn}

\begin{document}

\title{A confining quark model suggestion against $D_s^*(2317)$ and 
$D_s^*(2460)$ as chiral partners of standard $D_s$}
\author{P. Bicudo}
\affiliation{Dep. F\'{\i}sica and CFTP, Instituto Superior T\'ecnico,
Av. Rovisco Pais, 1049-001 Lisboa, Portugal}
\begin{abstract}
This paper presents the first study of mesons with a quark and an antiquark 
with different and finite masses, in a simple confining and chiral 
invariant quark-antiquark interaction, leading to spontaneous chiral 
symmetry breaking and to constituent quarks.
In the false chiral invariant vacuum, the chiral partners are degenerate, 
and tachyons occur in the light-light spectrum. In the true vacuum, 
most of the standard non-relativistic quark model spectra should be 
recovered except for the pion and other particular constraints. 
The calibration problem of chiral quark models is also addressed here.
The detailed inspection of the different contributions to the $D$ and $D_s$ masses 
suggests that the challenging recently observed  $D_s^*(2317)$ and $D_s^*(2460)$ 
mesons might not fit as global chirally rotated quark-antiquark $D_s$ mesons. 
\end{abstract}

\maketitle

\section{Introduction}

Recently, the discovery of the new $D_s^*(2317)$ and $D_s^*(2460)$
\cite{Babar,Cleo,Belle}
revived the interest in chiral partners. 
Ten years before the discovery, Nowak, Rho, Zahed,
Bardeen and Hill
\cite{Nowak, Bardeen} 
already predicted that the standard pseudoscalar meson $D_s(1968)$ 
and the standard vector meson $D_s^*(2112)$ would have two chiral partners, 
respectively a scalar and an axialvector, with masses compatible 
with the $D_s^*(2317)$ and $D_s^*(2460)$.
Indeed, mesons can be arranged in parity multiplets. In the false 
chiral invariant vacuum, the scalars and pseudoscalars, or the 
vectors and axialvectors, are degenerate. In the true, chiral 
symmetry breaking vacuum, their masses split and the important
question is, does this splitting explain the new $D_s$ resonances?

The revived conjecture of chiral partnership 
\cite{Nowak2,Bardeen2}
respectively between the $D_s^*(2317)$ and $D_s^*(2460)$, 
and the standard quark-antiquark mesons $D_s(1968)$ and $D_s^*(2112)$, 
is quite important because neither the quark model nor quenched lattice 
QCD are able to describe the $D_s^*(2317)$ and $D_s^*(2460)$ as standard 
quark-antiquark mesons. The new $D_s^*$ do not fit in the spectrum of 
standard quark-antiquark mesons, which is governed by the quark constituent 
masses and by a confining potential, together with well known hyperfine, 
spin-orbit and tensor potentials
\cite{Godfrey}, 
see Table \ref{algebraic}. 
Quenched lattice QCD, which only accesses the quark-antiquark spectrum,
confirms that these $D_s^*$ masses are too light for standard $q \bar q$ mesons
\cite{Bali,Dougall}.

Notice however that a calibration problem remains in all chiral 
computations of the hadron spectrum. For instance the first
models of chiral symmetry, like the $\sigma$ model of
Gell-Mann and Levy
\cite{Gell-Mann},
or the Nambu and Jona-Lasinio model
\cite{Nambu}, 
were only very accurate for the 
groundstate pseudoscalar mesons, because they did not address 
confinement. The ideal chiral framework should access the full 
phenomenology of the meson spectrum. 
I submit that this ideal framework is already under development. 
When the quarks were discovered, the confining quark model was
calibrated with correct confining and spin dependent potentials.
The first matrix elements of the spin-tensor potentials are
shown in Table \ref{algebraic}.
However it was realized that the main difficulty of the confining 
quark model consisted in understanding the low pion mass. 
But Nambu and Jona-Lasinio
\cite{Nambu}
had already shown that the spontaneous dynamical breaking of
global chiral symmetry provides a mechanism for the generation
of the constituent fermion mass and for the almost vanishing
mass of the pion. This mechanism was extended to the confining quark model 
by le Yaouanc, Oliver, Ono, P\`ene and Raynal with the Salpeter equations
in Dirac structure 
\cite{Yaouanc}
and by PB and Ribeiro with the equivalent Salpeter equations in a form
\cite{Bicudo_thesis}
identical to the Random Phase Approximation (RPA) equations
of Llanes-Estrada and Cotanch
\cite{Llanes-Estrada_thesis}.
Moreover, these chiral quark models also comply with the PCAC theorems, 
say the Gell-Mann Oakes and Renner relation
\cite{Yaouanc,Bicudo_scapuz},
the Adler Zero
\cite{Bicudo_PCAC,weall_pipi,Bicudo_piN},
the Goldberger-Treiman Relation 
\cite{Delbourgo,Bicudo_PCAC},
or the Weinberg Theorem
\cite{weall_pipi,Bicudo_PCAC,Llanes-Estrada_l1l2}.
However the correct fit of the hadronic spectra remains to
be fully addressed for confining and chiral invariant quark-antiquark
interactions. Nevertheless I submit that a confining chiral quark model 
with the correct spin-tensor potentials should eventually reproduce the 
full spectrum of hadrons, including heavy-light systems
\cite{Bicudo_thesis,Bardeen}.

For clarity, I now produce for the first time the full mesonic
spin-tensor potentials of a confining and chiral invariant quark model,
for a quark and an antiquark with different and finite masses. 
This is applied to study the $D$ and $D_s$ meson families, with
a quark $u, d$ or $s$ really lighter than the scale of QCD, and an
antiquark $c$ much heavier than the scale of QCD. The boundstate equations
are exactly solved to study chiral partners in the true vacuum and in
the limits of light or heavy quarks. This can be accomplished in the framework
of the simplest confining and chiral invariant quark model
\cite{Yaouanc,Bicudo_thesis,Bicudo_scapuz}.
The hamiltonian can be approximately derived from QCD, 
\begin{eqnarray}
&&H=\int\, d^3x \left[ \psi^{\dag}( x) \;(m_0\beta -i{\vec{\alpha}
\cdot \vec{\nabla}} )\;\psi( x)\;+
{ 1\over 2} g^2 \int d^4y\, \
\right.
\nonumber \\
&&
\overline{\psi}( x)
\gamma^\mu{\lambda^a \over 2}\psi ( x)  
\langle A_\mu^a(x) A_\nu^b(y) \rangle
\;\overline{\psi}( y)
\gamma^\nu{\lambda^b \over 2}
 \psi( y)  \ + \ \cdots
\label{hamilt}
\end{eqnarray}
up to the first cumulant order, of two gluons
\cite{Bicudo_hvlt,Dosch,Nefediev},
which can be evaluated in the modified coordinate gauge,
\begin{equation}
g^2 \langle A_\mu^a(x) A_\nu^b(y) \rangle
\simeq-{3 \over 4} \delta_{ab} g_{\mu 0} 
g_{\nu 0}
\left[K_0^3({\bf x}-{\bf y})^2-U\right]
\label{potential}
\end{equation}
and this is a simple density-density harmonic effective 
confining interaction. $m_0$ is the current mass of the quark,
and $K_0\simeq$ 0.3 to 0.4 GeV is the only physical scale in the interaction.
Like QCD, this model has only one scale in the interaction. 
The infrared constant $U$ confines the quarks but the 
meson spectrum is completely insensitive to it.

In Section II, starting from the confining and chiral invariant potential,
the mass and boundstate equations are derived for a quark and an antiquark
with different masses. In particular the spin-tensor potentials are studied in detail.
In Section III, the boundstate equations are applied to the $D$ and $D_s$ families. 
An interpolation from the ideal heavy-light limit in the false chiral
invariant vacuum, to the true symmetry breaking vacuum, and to finite current
quark masses is inspected in detail. In Section IV, I present the conclusion on 
the new $D_s^*(2317)$ and $D_s^*(2460)$ and on the calibration of confining and 
chiral invariant quark potentials.

%
%
\begin{table}[t]
\caption{\label{algebraic} 
Matrix elements of the spin-dependent potentials
}
\begin{ruledtabular}
\begin{tabular}{c|ccccc}
$^{2S+1}L_J$					& $\delta_{{\bf S}_q,{\bf S}_{\bar q}}$ 
									& $ {\bf S}_q \hspace{-.075 cm} 
									\cdot \hspace{-.075 cm} {\bf S}_{\bar q}$ 
											& $({\bf S}_q + {\bf S}_{\bar q}) \hspace{-.075 cm} 
											\cdot \hspace{-.075 cm} {\bf L}$
													& $({\bf S}_q - {\bf S}_{\bar q})\hspace{-.075 cm} 
													\cdot \hspace{-.075 cm} {\bf L}$
														& tensor		\\
$^1S_0$ 						&1	&-3/4	&0		&0	&0 				\\
$^3P_0$ 						&1	&1/4	&-2		&0	&-1/3 			\\
$^3S_1$ 						&1	&1/4	&0		&0	&0 				\\
$^3D_1$ 						&1	&1/4	&-3		&0	&-1/6 			\\
$^3S_1\leftrightarrow {}^3D_1$	&0	&0		&0		&0	&$\sqrt{2}$/6 	\\
$^1P_1$ 						&1	&-3/4	&0		&0	&0 				\\
$^3P_1$ 						&1	&1/4	&-1		&0	&1/6 			\\
$^1P_1\leftrightarrow {}^3P_1$	&0	&0		&0		&$\sqrt{2}$	&0			 	\\
\end{tabular}
\end{ruledtabular}
\end{table}

\section{Mass gap and boundstate equations}

\begin{figure}[t]
\caption{
The constituent quark masses $m_c(k)$, solutions of the mass gap equation,
for different current quark masses $m_0$. 
}\label{mass solution}
\includegraphics[width=0.90\columnwidth]{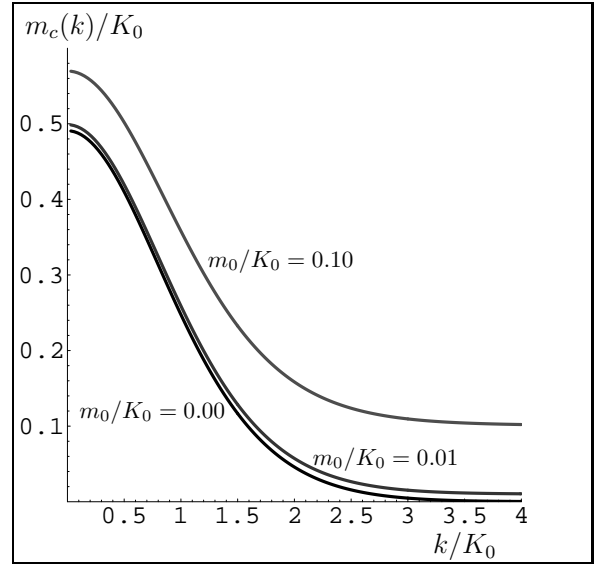}
\end{figure}

The relativistic invariant Dirac-Feynman propagators
\cite{Yaouanc}, 
can be decomposed in the quark and antiquark Bethe-Goldstone 
propagators
\cite{Bicudo_scapuz},
close to the formalism of non-relativistic quark models,
\FL
\begin{eqnarray}
{\cal S}_{Dirac}(k_0,\vec{k})
&=& {i \over \not k -m +i \epsilon}
\nonumber \\
&=& {i \over k_0 -E(k) +i \epsilon} \
\sum_su_su^{\dagger}_s \beta
\nonumber \\
&& - {i \over -k_0 -E(k) +i \epsilon} \
\sum_sv_sv^{\dagger}_s \beta \ ,
\nonumber \\
u_s({\bf k})&=& \left[
\sqrt{ 1+S \over 2} + \sqrt{1-S \over 2} \widehat k \cdot \vec \sigma \gamma_5
\right]u_s(0)  \ ,
\nonumber \\
v_s({\bf k})&=& \left[
\sqrt{ 1+S \over 2} - \sqrt{1-S \over 2} \widehat k \cdot \vec \sigma \gamma_5
\right]v_s(0)  \ ,
\nonumber \\
&=& -i \sigma_2 \gamma_5 u_s^*({\bf k}) \ ,
\label{propagators}
\end{eqnarray}
where $S=\sin(\varphi)={m_c\over \sqrt{k^2+m_c^2}} \ , 
\ C=\cos(\varphi)={k\over \sqrt{k^2+m_c^2}}$ and $\varphi$ is a chiral angle.
In the non condensed vacuum, $\varphi$ is equal to $\arctan{m_0 \over k}$,
but $\varphi$ is not determined from the onset when chiral symmetry breaking
occurs.
In the physical vacuum, the constituent quark mass $m_c(k)$, or the
chiral angle $\varphi(k)=\arctan{m_c(k) \over k}$, is a variational function
which is determined by the mass gap equation. Examples of solutions,
for different light current quark masses $m_0$, 
are depicted in Fig. \ref{mass solution}.

%

Then there are three equivalent methods to find 
the true and stable vacuum, where constituent quarks acquire 
the constituent mass.
One method consists in assuming a quark-antiquark $^3P_0$ 
condensed vacuum, and in minimizing the vacuum energy density. 
A second method consists in rotating the quark and antiquark 
fields with a Bogoliubov-Valatin canonical transformation 
to diagonalize the terms in the hamiltonian with two   
quark or antiquark second quantized fields. 
A third method consists in solving the Schwinger-Dyson 
equations for the propagators. Any of these methods
lead to the same mass gap equation and to the quark 
dispersion relation. Here I replace the propagator
of eq. (\ref{propagators}) in the Schwinger-Dyson equation, 
\begin{eqnarray}
\label{2 eqs}
&&0 = u_s^\dagger(k) \left\{k \widehat k \cdot \vec \alpha + m_0 \beta
-\int {d w' \over 2 \pi} {d^3k' \over (2\pi)^3}
i V(k-k') \right.
\nonumber \\
&&\left. \sum_{s'} \left[ { u(k')_{s'}u^{\dagger}(k')_{s'} 
 \over w'-E(k') +i\epsilon}
-{ v(k')_{s'}v^{\dagger}(k')_{s'} 
  \over -w'-E(k')+i\epsilon} \right]
\right\} v_{s''}(k) \  \
\nonumber \\
&&E(k) = u_s^\dagger(k) \left\{k \widehat k \cdot \vec \alpha + m_0 \beta
-\int {d w' \over 2 \pi} {d^3k' \over (2\pi)^3}
i V(k-k')  \right.
\nonumber \\
&&\left. \sum_{s'} \left[ { u(k')_{s'}u^{\dagger}(k')_{s'} 
 \over w'-E(k') +i\epsilon}
-{   v(k')_{s'}v^{\dagger}(k')_{s'}  
 \over -w'-E(k')+i\epsilon} \right]
\right\} u_s(k),
\end{eqnarray}
where, with the simple density-density harmonic interaction
\cite{Yaouanc}, the integral of the potential is a laplacian 
and the mass gap equation and the quark energy are finally,
\begin{eqnarray}
\label{mass gap}
\Delta \varphi(k)  &=& 2 k S(k) -2 m_0 C(k) - { 2 S(k) C(k) \over k^2 }  
\\ \nonumber 
E(k)&=& k C(k) + m_0 S(k) - { {\varphi'(k) }^2 \over 2 } - { C(k)^2 \over k^2 } 
+{ U \over 2} \ .
\end{eqnarray}
Numerically, this equation is a non-linear ordinary differential
equation. It can be solved with the Runge-Kutta and shooting method.
Examples of solutions for the current quark mass $m_c(k)= k \tan \varphi$, 
for different current quark masses $m_0$, 
are depicted in Fig. \ref{mass solution}.

%
%
\begin{table}[t]
\caption{\label{spin dependent} 
The positive and negative energy spin-independent, spin-spin, spin-orbit and 
tensor potentials are shown, for the simple density-density harmonic model
of eq. (\ref{potential}).
$\varphi'(k)$, ${\cal C}(k)$ and ${\cal G}(k)= 1 - S(k) $ are all functions of the constituent 
quark(antiquark) mass.
}
\begin{ruledtabular}
\begin{tabular}{c|c}
& $V^{++}=V^{--}$  \\ \hline
spin-indep. & $- {d^2 \over dk^2 } + { {\bf L}^2 \over k^2 } + 
{1 \over 4} \left( {\varphi'_q}^2 + {\varphi'_{\bar q}}^2 \right) 
+ {  1 \over k^2} \left( {\cal G}_q +{\cal G}_{\bar q}  \right) -U $  \\ 
spin-spin & $ {4 \over 3 k^2} {\cal G}_q {\cal G}_{\bar q} {\bf S}_q \cdot {\bf S}_{\bar q} $  \\ 
spin-orbit & $ {1 \over  k^2} \left[ \left( {\cal G}_q + 
{\cal G}_{\bar q} \right) \left( {\bf S}_q +{\bf S}_{\bar q}\right) 
+\left( {\cal G}_q - {\cal G}_{\bar q} \right) \left( {\bf S}_q -{\bf S}_{\bar q}\right)  \right]
\cdot {\bf L} $  \\ 
tensor & $ -{2 \over  k^2} {\cal G}_q {\cal G}_{\bar q} 
\left[ ({\bf S}_q \cdot \hat k ) ({\bf S}_{\bar q} \cdot \hat k )
-{1 \over 3} {\bf S}_q \cdot {\bf S}_{\bar q} \right] $ \\ \hline
& $V^{+-}=V^{-+}$ \\ \hline
spin-indep. & $0$  \\ 
spin-spin & $ -{4 \over 3} \left[ {1\over 2} {\varphi'_q} {\varphi'_{\bar q}} + 
{1\over k^2} {\cal C}_q {\cal C}_{\bar q}  \right]
{\bf S}_q \cdot {\bf S}_{\bar q} $  \\ 
spin-orbit & $0$  \\ 
tensor & $ \left[ -2 {\varphi'_q} {\varphi'_{\bar q}} + 
{2\over k^2} {\cal C}_q {\cal C}_{\bar q}  \right]
\left[ ({\bf S}_q \cdot \hat k ) ({\bf S}_{\bar q} \cdot \hat k )
-{1 \over 3} {\bf S}_q \cdot {\bf S}_{\bar q} \right] $
\end{tabular}
\end{ruledtabular}
\end{table}

The Salpeter-RPA equations for a meson (a colour singlet
quark-antiquark bound state) can be derived from the Lippman-Schwinger
equations for a quark and an antiquark, or replacing the propagator
of eq. (\ref{propagators}) in the Bethe-Salpeter equation. In either way, one gets
\cite{Bicudo_scapuz}
\FL
\begin{eqnarray}
\label{homo sal}
\phi^+(k,P) &=& { u^\dagger(k_1) \chi(k,P)  v(k_2) 
\over +M(P)-E(k_1)-E(k_2) }
\nonumber \\
{\phi^-}^t(k,P) &=& { v^\dagger(k_1) \chi(k,P) u(k_2)
\over -M(P)-E(k_1)-E(k_2)}
\nonumber \\
\chi(k,P) &=&
\int {d^3k' \over (2\pi)^3} V(k-k') \left[ 
u(k'_1)\phi^+(k',P)v^\dagger(k'_2) \right.
\nonumber \\
&&\left. +v(k'_1){\phi^-}^t(k',P) u^\dagger(k'_2)\right] 
\end{eqnarray}
where $k_1=k+{P \over 2} \ , \ k_2=k-{P \over 2}$ and $P$ is
the total momentum of the meson.
Notice that, solving for $\chi$, one gets the Salpeter equations of Yaouanc et al.
\cite{Yaouanc}.

The Salpeter-RPA equations of PB et al. 
\cite{Bicudo_thesis}
and of Llanes-Estrada et al. 
\cite{Llanes-Estrada_thesis}
are obtained deriving the equation for the positive energy wavefunction
$\phi^+$ and for the negative energy wavefunction $\phi^-$. The 
relativistic equal time equations have the double of coupled
equations than the Schr\"odinger equation, although in many cases the
negative energy components can be quite small. This results in four potentials 
$V^{\alpha \beta}$ respectively coupling $\nu^\alpha=r \phi^\alpha$ to 
$\nu^\beta$. The Pauli $\vec \sigma$ matrices in the spinors of eq. (\ref{propagators}) 
produce the spin-dependent
\cite{Bicudo_baryon} 
potentials of Table \ref{spin dependent}. 

Notice that both the pseudoscalar and scalar equations
have a system with two equations. This is the minimal number of relativistic 
equal time equations. However the spin-dependent interactions 
couple an extra pair of equations both in the vector and axialvector channels.
While the coupling of the s-wave and the d-wave are standard in vectors, the coupling 
of the spin-singlet and spin-triplet in axialvectors only occurs if the quark and antiquark 
masses are different, say in heavy-light systems. 
I now combine the algebraic matrix elements of Table \ref{algebraic}
with the spin-dependent potentials of Table \ref{spin dependent},
to derive the full Salpeter-RPA radial boundstate
equations (where the infrared $U$ is dropped from now on). 
I get the $J^P=0^{-}$,  $^1 S_0$ pseudoscalar ($P$) equations,

\onecolumngrid
\begin{equation}
\label{pseudoscalar}
\left\{ \left( -{d^2 \over d k ^2} +E_q(k) +E_{\bar q}(k)  
+ { {\varphi'_q}^2 +{\varphi'_{\bar q}}^2 \over 4} + {1-S_q S_{ \bar q} \over k^2}  \right)
\left[ \begin{array}{cc}
1 & 0 \\ 0 & 1 \end{array} \right]
+
\left( {\varphi'_q \varphi'_{\bar q} \over 2} + {C_q C_{\bar q} \over k^2 } \right)
\left[ \begin{array}{cc}
0 & 1 \\ 1 & 0 \end{array} \right]
-M 
\left[ \begin{array}{cc}
1 & 0 \\ 0 & -1 \end{array} \right]
\right\}
\left( \begin{array}{c} \nu_{^1S_0}^+(k) \\ \nu_{^1S_0}^-(k) \end{array} \right) = 0
\  ,
\end{equation}
the $J^P=0^{+}$, $^3 P_0$ scalar ($S$) equations,
\begin{equation}
\label{scalar}
\left\{ \left( -{d^2 \over d k ^2} +E_q(k) +E_{\bar q}(k)  
+ { {\varphi'_q}^2 +{\varphi'_{\bar q}}^2 \over 4} + {1+S_q S_{ \bar q} \over k^2}  \right)
\left[ \begin{array}{cc}
1 & 0 \\ 0 & 1 \end{array} \right]
+
\left( {\varphi'_q \varphi'_{\bar q} \over 2} - {C_q C_{\bar q} \over k^2 } \right)
\left[ \begin{array}{cc}
0 & 1 \\ 1 & 0 \end{array} \right]
-M 
\left[ \begin{array}{cc}
1 & 0 \\ 0 & -1 \end{array} \right]
\right\}
\left( \begin{array}{c} \nu_{^3P_0}^+(k) \\ \nu_{^3P_0}^-(k) \end{array} \right) = 0
\  .
\end{equation}
the $J^P=1^{-}$,  coupled $^3 S_1$ and $^3D_1$ vector ($V$ and $V^*$) equations ,
\begin{eqnarray}
\label{vector}
\left\{ 
\left( -{d^2 \over d k ^2} +E_q(k) +E_{\bar q}(k)  
+ { {\varphi'_q}^2 +{\varphi'_{\bar q}}^2 \over 4} 
+ {7-4S_q -4S_{ \bar q}+S_q S_{ \bar q}\over 3 k^2} \right)
\left[ \begin{array}{cccc}
1 & 0 & 0 & 0 \\ 
0 & 1 & 0 & 0 \\ 
0 & 0 & 0 & 0 \\ 
0 & 0 & 0 & 0 
\end{array} \right]
+
\left( -{\varphi'_q \varphi'_{\bar q} \over 6} 
-{C_q C_{\bar q} \over 3k^2 } \right)
\left[ \begin{array}{cccc}
0 & 1 & 0 & 0 \\ 
1 & 0 & 0 & 0 \\
0 & 0 & 0 & 0 \\
0 & 0 & 0 & 0
\end{array} \right]
\right. &&
\\ \nonumber 
+
\left( -{d^2 \over d k ^2} +E_q(k) +E_{\bar q}(k)  
+ { {\varphi'_q}^2 +{\varphi'_{\bar q}}^2 \over 4} 
+ {8+4S_q +4S_{ \bar q}+2S_q S_{ \bar q}\over 3 k^2} \right)
\left[ \begin{array}{cccc}
0 & 0 & 0 & 0 \\ 

0 & 0 & 0 & 0 \\ 
0 & 0 & 1 & 0 \\ 
0 & 0 & 0 & 1 
\end{array} \right]
+
\left(  {\varphi'_q \varphi'_{\bar q} \over 6} 
-{ 2 C_q C_{\bar q} \over 3k^2 } \right)
\left[ \begin{array}{cccc}
0 & 0 & 0 & 0 \\ 
0 & 0 & 0 & 0 \\
0 & 0 & 0 & 1 \\
0 & 0 & 1 & 0
\end{array} \right] 
&&
\\ \nonumber 
\left.
- 
{ \left(1-S_q\right) \left(1-S_{ \bar q}\right)\over 3 k^2 } 
\left[ \begin{array}{cccc}
0 & 0 & \sqrt{2} & 0 \\ 
0 & 0 & 0 & \sqrt{2} \\ 
\sqrt{2} & 0 & 0 & 0 \\ 
0 & \sqrt{2} & 0 & 0 

\end{array} \right]
-\left(  {\varphi'_q \varphi'_{\bar q} \over 3} 
- {C_q C_{\bar q} \over 3k^2 } \right) 
\left[ \begin{array}{cccc}
0 & 0 & 0 & \sqrt{2} \\ 
0 & 0 & \sqrt{2} & 0 \\
0 & \sqrt{2} & 0 & 0 \\
\sqrt{2} & 0 & 0 & 0
\end{array} \right]
-M 
\left[ \begin{array}{cccc}
1 & 0 & 0 & 0 \\ 
0 & -1 & 0 & 0 \\ 
0 & 0 & 1 & 0 \\ 
0 & 0 & 0 & -1 
\end{array} \right]
\right\} 
\left( \begin{array}{c} \nu_{^3S_1}^+(k) \\ \nu_{^3S_1}^-(k) \\ \nu_{^3D_1}^+(k) \\ \nu_{^3D_1}^-(k) 
\end{array} \right) 
& =& 0
\  ,
\end{eqnarray}
the $J^P=1^{+}$, coupled $^1P_1$ and $^3P_1$ axialvector ($A$ and $A^*$) equations 
\begin{eqnarray}
\label{axialvector}
\left\{ \left( -{d^2 \over d k ^2} +E_q(k) +E_{\bar q}(k)  
+ { {\varphi'_q}^2 +{\varphi'_{\bar q}}^2 \over 4} 
+ {3-S_q S_{ \bar q} \over k^2}  \right)
\left[ \begin{array}{cccc}
1 & 0 & 0 & 0 \\ 
0 & 1 & 0 & 0 \\ 
0 & 0 & 0 & 0 \\ 
0 & 0 & 0 & 0 
\end{array} \right]
+
\left( {\varphi'_q \varphi'_{\bar q} \over 2} + {C_q C_{\bar q} \over k^2 } \right)
\left[ \begin{array}{cccc}
0 & 1 & 0 & 0 \\ 
1 & 0 & 0 & 0 \\
0 & 0 & 0 & 0 \\
0 & 0 & 0 & 0
\end{array} \right]
\right.
&&
\\ \nonumber 
\left( -{d^2 \over d k ^2} +E_q(k) +E_{\bar q}(k)  
+ { {\varphi'_q}^2 +{\varphi'_{\bar q}}^2 \over 4} 
+{2 \over k^2} \right)
\left[ \begin{array}{cccc}
0 & 0 & 0 & 0 \\ 
0 & 0 & 0 & 0 \\ 
0 & 0 & 1 & 0 \\ 
0 & 0 & 0 & 1 
\end{array} \right]
+
\left( - {\varphi'_q \varphi'_{\bar q} \over 2} \right)
\left[ \begin{array}{cccc}
0 & 0 & 0 & 0 \\ 
0 & 0 & 0 & 0 \\
0 & 0 & 0 & 1 \\
0 & 0 & 1 & 0
\end{array} \right] 
&&
\\ \nonumber 
\left.
+
{ S_q - S_{\bar q}\over  k^2 } 
\left[ \begin{array}{cccc}
0 & 0 & \sqrt{2} & 0 \\ 
0 & 0 & 0 & \sqrt{2} \\ 
\sqrt{2} & 0 & 0 & 0 \\ 
0 & \sqrt{2} & 0 & 0 
\end{array} \right]
-M 
\left[ \begin{array}{cccc}
1 & 0 & 0 & 0 \\ 
0 & -1 & 0 & 0 \\ 
0 & 0 & 1 & 0 \\ 
0 & 0 & 0 & -1 
\end{array} \right]
\right\} 
\left( \begin{array}{c} \nu_{^1P_1}^+(k) \\ \nu_{^1P_1}^-(k) \\ \nu_{^3P_1}^+(k) \\ \nu_{^3P_1}^-(k) 
\end{array} \right) 
& =& 0
\  ,
\end{eqnarray}

\twocolumngrid

In the light-light limit of $m_q =m_{\bar q} \rightarrow 0$ and $\varphi \rightarrow 0$, 
it is clear that eq. (\ref{pseudoscalar}) and eq. (\ref{scalar}) become identical. They
also possess takyonic solutions 
\cite{Yaouanc}. 
In the same limit, eq. (\ref{vector}) can be block diagonalized
\cite{Yaouanc}, 
and each block, with mixed s-wave and d-wave, is identical one of the two 
independent blocks of eq. (\ref{axialvector}). This checks that the chiral partners 
$P$-$S$ and $V,V^*$-$A,A^*$ are degenerate in the false chiral symmetric vacuum.

Another interesting case is the heavy-light case where, say, the antiquark has a mass 
$m_{\bar q}\simeq {m_0}_{\bar q} >> K_0$, there are no Tachyons, and the negative energy 
components nearly vanish, like in non-relativistic quark models. 
In the infinite $m_{\bar q}$ limit, $S_{\bar q} \rightarrow 1$, and the antiquark spin 
is irrelevant, see Table \ref{spin dependent}, complying with the 
Isgur-Wise heavy-quark symmetry
\cite{Isgur}.

For the numerical solution, I change the sign of the 
second and fourth lines in eqs (\ref{pseudoscalar}) to 
(\ref{axialvector}) and, replacing the derivatives of
the wave-functions by finite difference matrices,
the equations become simple eigenvalue equations. 

\section{Results for the $D_s$ and $D$ mesons}

I now compute in detail $D$ and $D_s$ masses, relevant 
to the conjecture of chiral partnership respectively between 
the scalar meson $D_s^*(2317)$ and the axialvector meson
$D_s^*(2460)$, and the standard quark-antiquark pseudoscalar 
meson $D_s(1968)$ and vector meson $D_s^*(2112)$. It is 
convenient to start from the chiral invariant false vacuum 
where, in  the ideal heavy-light limit of a massless quark and
and infinitely massive antiquark, the groundstate
pseudoscalar is degenerate with the groundstate 
scalar, and the groundstate vector is degenerate with
the groundstate axialvector.

Then, interpolating from this ideal limit to the
actual constituent masses of the light quark
and of the heavy antiquark, the mass splittings 
between the $D_s^*(2317)$ and the $D_s(1968)$ 
and between the $D_s^*(2460)$ and the $D_s^*(2112)$
can be computed. To inspect in detail the contributions
to the mass splittings, it is important to decompose
this chiral interpolation in three different steps.  

In the first step the current quark masses
are in the ideal chiral limit of ${m_0}_q=0$ and in the
ideal Isgur-Wise limit of ${m_0}_{\bar q}=\infty$,
and I interpolate the quark mass $m_q$ from
$0$, corresponding to the false chiral invariant vacuum, to
the actual constituent quark mass ${m_c}_q=0$ solution of the
mass gap equation (\ref{mass gap}). 

In the second step the current mass ${m_0}_{\bar q}$ of the heavy 
antiquark is interpolated from the the ideal Isgur-Wise limit of 
${m_0}_{\bar q}=\infty$, to its actual value of the order of
$m0_{\bar q}\simeq 5 K_0$, fitted in the $J/\Psi$ spectrum. 
Notice that in the case of heavy quarks or antiquarks, the 
constituent quark mass is identical to the current quark mass, 
the mass gap equation (\ref{mass gap}) essentially does not 
change the heavy quark masses.

I leave for the third and final step the interpolation
of the current quark mass ${m_0}_q$ from the ideal chiral limit
to the actual values of the order of ${m_0}_{ q}\simeq 0.01 K_0$
for the $u$ and $d$ and of ${m_0}_{ q}\simeq 0.1 K_0$ for the
$s$ quark. In chiral models the current masses of light quarks
are model dependent. Although these current masses $m_0$ are 
smaller than the ones used, say, in Chiral Lagrangians, in this 
model these current quark masses are the ones that lead to the 
correct experimental masses of the light-light $\pi$ and $K$ mesons.
Therefore our $m_0$ are not free parameters.  

The results are respectively inspected
in Subsection A, in Subsection B and in Subsection C, 
and are respectively depicted in Fig. \ref{heavy-light_1},
Fig. \ref{heavy-light_2} and Fig. \ref{heavy-light_3}. 
Notice that, if the three figures are placed side by
side, the interpolations of the studied mesons
exactly match. At the end of the three interpolations
the $D$ and $D_s$ spectra is computed.

\subsection{From the chiral invariant to the true vacua}

In the first step the quark current mass
is in the ideal chiral limit of ${m_0}_q=0$, 
the antiquark current mass is in the
ideal Isgur-Wise limit of ${m_0}_{\bar q}=\infty$,
and I interpolate the quark mass $m_q$ from
$0$, corresponding to the false chiral invariant vacuum, to
the actual constituent quark mass ${m_c}_q=0$ solution of the
mass gap equation (\ref{mass gap}). 

When the antiquark has an infinite mass, all terms depending
on its spin vanish. In table \ref{spin dependent} it is clear
that a quark, or antiquark spin always comes with the factor,
\begin{equation}
{\cal G}(k)= 1 - {m_c \over \sqrt{k^2+ {m_c}^2}} \ .
\end{equation}
Thus ${\cal G}(k)$ is maximal and equal to 1 when $m_c=0$
and ${\cal G}(k)$ is minimal and equal to 0 when $m_c=\infty$.
Because the spin of the heavy antiquark is irrelevant, the masses
of the groundstate pseudoscalar $P$ and vector $V$ are degenerate, 
and the masses of the groundstate scalar $S$ and axialvector $A$ 
are also degenerate. Thus I get, in the present limit
\begin{equation}
M_A-M_V = M_S-M_P \ .
\label{chiral relation exact}
\end{equation}

%
\begin{figure}[t]
\caption{
Heavy-light meson masses. Here ${m_0}_{\bar q}=\infty$
and ${m_0}_q=0$. 
The light constituent quark mass is interpolated from the zero mass of 
the chiral invariant false vacuum to the solution $m_c$ of the 
mass gap equation in the true vacuum.  
}\label{heavy-light_1}
\includegraphics[width=1.0\columnwidth]{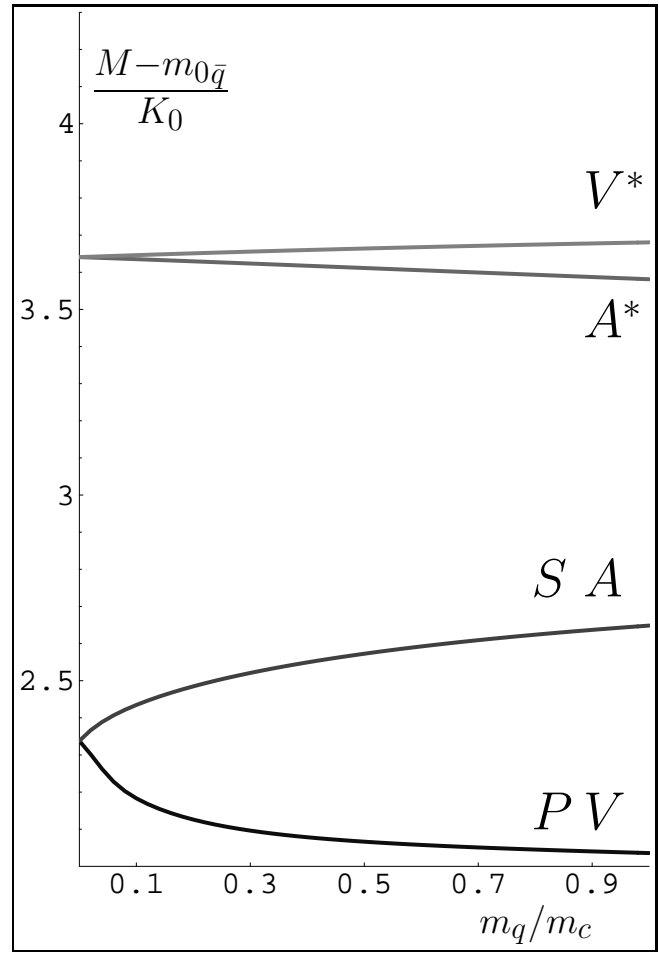}
\end{figure}

Moreover, the only spin-dependent term that does not vanish in this 
case is the spin-orbit term 
${ 2 \over k^2} {\cal G}_q {\bf S}_q \cdot {\bf L}$. 
Thus the mass splittings of eq . (\ref{chiral relation exact}) 
measure the angular repulsive barrier and the spin-orbit term.

Now, in the chiral invariant false vacuum 
the spin-orbit term simplifies to 
${ 2 {\bf S}_q \cdot {\bf L}  \over k^2} $. In this
case the spin-orbit term is able to fully compete with the angular
repulsive barrier $l(l+1) \over  k^2$, and the spectrum only
depends on the total angular momentum ${\bf J}={\bf L} + {\bf S}_q $
of the light quark,
\begin{equation}
{1 \over k^2}{\bf L}^2 + {2 \over k^2} {\bf S}_q \cdot {\bf L} = 
{1 \over k^2}\left({\bf J}^2 - {{\bf S}_q}^2 \right)  \ ,
\end{equation}
independently of $\bf L$.
Thus, in the chiral invariant false vacuum, chiral symmetry
induces an extra degeneracy in the states,
$P, \, V , \, S \, A$, in the states $A^* , \, V^*$ and so on.

In the true vacuum the quark mass is the finite constituent quark 
mass ${m_q}_c$, and this decreases the spin-orbit interaction
${ 2 \over k^2} {\cal G}_q {\bf S}_q \cdot {\bf L}$, which is
no longer able to cancel the mass splittings induced by the
angular repulsive barrier $l(l+1) \over  k^2$. In the limit
when this spin-orbit interaction vanishes, the splittings are
only due to the repulsive barrier.

The opposite limit of large spin-orbit may occur in the case of 
very large angular excitations
\cite{Yaouanc,Bicudo_hvlt,Malheiro,Kalashnikova},
leading to chiral doubles in the spectrum, even when the full
constituent mass is used.

Notice that this first step accounts for most of the splitting of
eq. (\ref{chiral relation exact}). In Fig. \ref{heavy-light_1}, 
this splitting is already of the order of 0.61 $K_0$. After the
three steps it will be of the order of 0.81 $K_0$.
This is smaller, but of a comparable order, than the typical 
scale of angular splittings of the hadronic spectra.

\subsection{From the heavy quark limit to the $c$ quark}

%
\begin{figure}[t]
\caption{
Heavy-light meson masses. Here, ${m_0}_q=0$ and $m_q(k)=m_c(k)$ remain
 in the chiral limit. The heavy antiquark 
masses decreases from the infinite limit of Isgur-Wise to the 
actual charm mass ${m_0}_{\bar q} \simeq 5 K_0$. 
}\label{heavy-light_2}
\includegraphics[width=1.0\columnwidth]{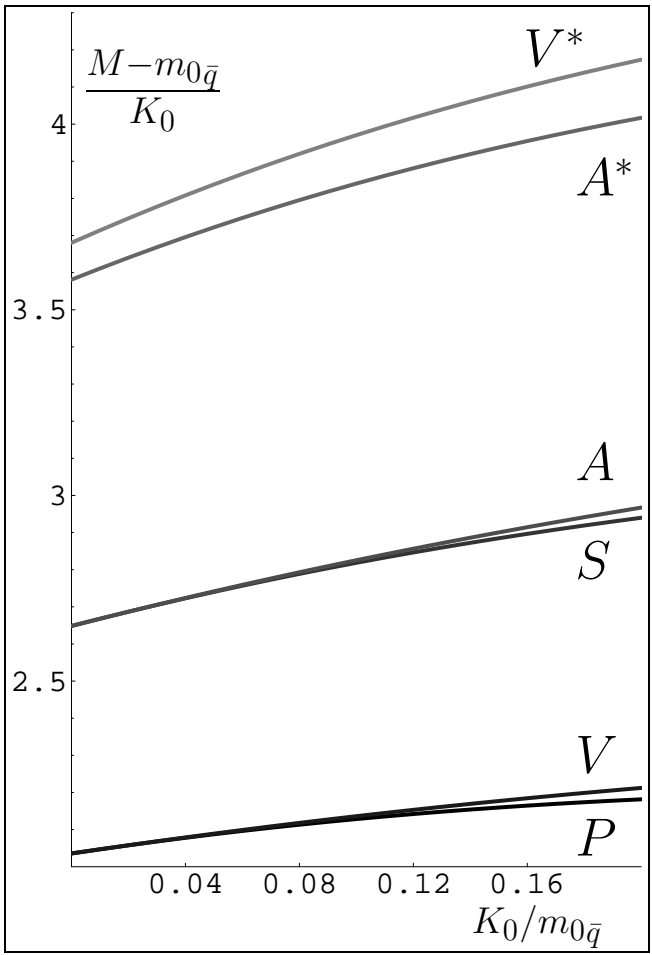}
\end{figure}

In the second step the current mass ${m_0}_{\bar q}$ of the heavy 
antiquark is interpolated from the ideal isgur-wise limit of 
${m_0}_{\bar q}=\infty$, to its actual value of the order of
${m_0}_{\bar q}\simeq 5 K_0$. Notice that in the case of heavy
quarks or antiquarks, the constituent quark mass is very close to 
the current quark mass. In this case, the mass gap equation only
changes the quark mass in a negligible way. Thus this also 
interpolates the constituent antiquark mass from 
$\infty$ to ${m_0}_{\bar q}\simeq 5 K_0$.

In this step the spin-spin and the tensor potentials no longer
vanish. In Fig. \ref{heavy-light_2}, these spin-dependent
potentials are able to split the masses of the pseudoscalar
and vector and the masses of the scalar and axialvector. It is
remarkable that these two mass splittings are almost identical,
\begin{equation}
M_V-M_P \simeq M_A-M_S \ ,
\label{chiral relation approximate}
\end{equation}
with a precision better than 1 per mil.
For this result both the spin-spin and tensor
interactions have to conspire with a beautiful precision.

Nevertheless these hyperfine and tensor splittings are too small. 
This happens because in this model the $\cal G$ function,
defined in Table \ref{spin dependent}, suffers from 
steep dependence on the quark mass
\begin{equation}
 \lim_{m \to \infty} {\cal G} = { k^2 \over 2 m_c^2 } 
\end{equation}
while it is well known from phenomenology that the spin-spin 
interaction dependence on the constituent quark masses
is much smoother. Thus the splittings in Fig. \ref{heavy-light_2}, 
are more than one order of magnitude smaller than the experimental 
splittings.

\subsection{From the chiral limit to the $u,d$ and $s$ quarks}

%
\begin{figure}[t]
\caption{
Heavy-light meson masses. Here ${m_0}_{\bar q}$  is the
charm mass, and the light quark current quark mass $m_0$ interpolates 
from the vanishing mass of the chiral limit, passes by the $u$ and $d$
current quark masses and ends up at the $s$ quark mass. 
}\label{heavy-light_3}
\includegraphics[width=1.00\columnwidth]{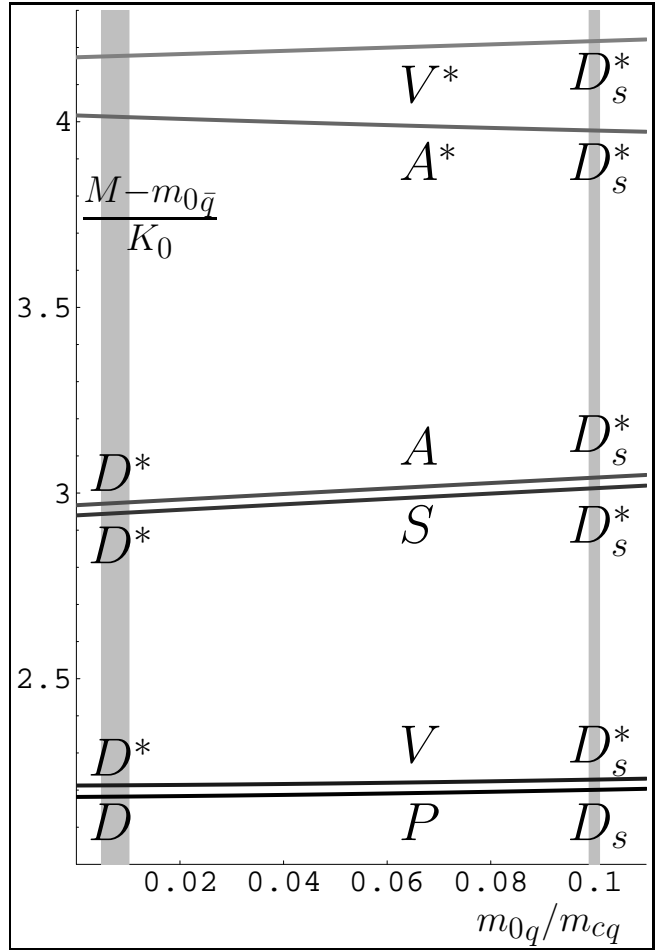}
\end{figure}

I leave for the third and final step the interpolation
of the current quark mass ${m_0}_q$ from the ideal chiral limit
to the actual values of the order of $m0_{ q}\simeq 0.01 K_0$
for the $u$ and $d$ and of $m0_{ q}\simeq 0.1 K_0$ for the
$s$ quark. In chiral models the current masses of light quarks
are model dependent. Although they are smaller than the ones used, say, 
in Chiral Lagrangians, in this model these current quark masses 
are the ones that lead to the correct experimental masses of the 
light-light $\pi$ and $K$ mesons.

Notice that interpolating from vanishing to finite current quark 
masses, in this model, essentially does not change the $M_V-M_P$
and $M_A-M_S$ splittings. Essentially the $M_S-M_V$ splitting is 
slightly increased and the $M_{A^*}-M_A$ splitting is slightly 
decreased.

\section{Conclusion}

For the first time a quark model with a chiral 
symmetric and confining interaction is applied to compute exactly
different $D$ and $D_s$ meson masses for finite $u, \, d, \, s$ and $c$ 
current quark masses. The different spin-tensor contributions to the
meson masses are also analyzed in detail. I now discuss the results
both qualitatively and quantitatively, and address the new
$D_s^*(2317)$ and $D_s^*(2460)$ resonances.

My quantitative conclusion is that chiral models have the same number of 
meson states in the spectrum as the normal quark model. The mass splittings 
can be related, as usual in quark models, to spin-tensor potentials. At the 
same token the spectrum complies with the chiral relations. For instance
the well known mass formula, first predicted by the heavy-light 
chiral papers
\cite{Nowak,Bardeen},
 \begin{equation}
M_A-M_V \simeq M_S-M_P
\label{chiral relation}
\end{equation}
is correct, in this model, up to the fourth decimal case.
It is quite remarkable that all the spin and angular momentum
tensor potentials precisely conspire to achieve this  
result. 
Therefore I confirm that eq.
(\ref{chiral relation})
must be correct for the standard quark-antiquark mesons 
$D_s(1968)$, the $D_s(2112)$  and for their scalar and axialvector chiral 
partners.
Notice however that a very similar pattern to the one of eq.
(\ref{chiral relation}) also occurs within the 
$D$ sector, see in Fig. \ref{heavy-light_1}, Fig. \ref{heavy-light_2} 
and Fig. \ref{heavy-light_3}. The similar pattern of the quark-antiquark,
or quenched, spectra for the $D$ and $D_s$ family is expected in confining 
quark models but here it is mentioned for the first time in a chiral
calculation.

Before the quantitative conclusion is presented, notice that,
quantitatively, all chiral models, including this simple 
density-density harmonic confining model of 
eq. (\ref{potential}), and the chiral models of Nowak, Rho and Zahed
of Bardeen and Hill, suffer from a calibration problem. 
The present model is confining, so it belongs to a class of models 
already able to fit the angular and radial excitations of the 
hadronic spectra. 
In this sense this constitutes and upgrade of the non-confining
$\sigma$-model
\cite{Gell-Mann}, 
of the Nambu and Jona-Lasinio Model
\cite{Nambu}
and of te related models of 
Nowak, Rho and Zahed and of Bardeen and Hill.
Nevertheless the spin-tensor interactions remain 
to be calibrated, and this is precisely addressed in this paper.
This calibration problem is equivalent to the 
problem of chiral symmetry with scalar confinement recently mentioned,
for instance, by Adler
\cite{Adler}.
Notice that the calibration problem of chiral quark models is quite
important. If this problem was solved, the confining quark
model would be further improved, both in accuracy because
the pion mass and other particular constrains like eq. (\ref{chiral relation})
would be correct, and in consistency because fewer parameters
would be needed to fit the hadron spectra.
But I submit that the under development chiral invariant 
quark models with a confining funnel interaction
\cite{Bicudo_KN,Llanes-Estrada_thesis} 
including a short range vector interaction
\cite{Llanes-Estrada_hyperfine,Bicudo_scapuz},
and a long range confining scalar interaction
\cite{Bicudo_scalar,Villate},
can be correctly calibrated.
Llanes-Estrada, Cotanch, Szczepaniak and Swanson showed that
the Coulomb potential is crucial to produce correct
hyperfine splittings both for light and heavy quark masses.
Possibly the scalar confining potential suggested by
PB and Marques would also suppress the spin-orbit interaction.
An important example is provided by quenched lattice QCD computations
with Ginsparg-Wilson or Staggered fermions, which reproduce 
the spectrum of quark-antiquark mesons, including the light pion mass.  
Then the correct implementation of chiral symmetry should not 
affect the broad picture of the quark model spectrum of 
$q \bar q$ mesons except for particular constraints like 
the low pion mass and eq. (\ref{chiral relation}). 

Quantitatively, our result for the splittings of eq. (\ref{chiral relation}),
depicted in Fig. \ref{heavy-light_3}, may be as large as 325 MeV, for the upper 
bound of 400 MeV for the potential strength $K_0$. This $M_S-M_P$ splitting
is the crucial one for the chiral $D_s$ conjecture (although I also mention
here the hyperfine splitting $M_V-M_P$).
Notice that this is close to the splitting of 350 MeV advocated by the conjecture 
of chiral partnership 
\cite{Nowak2,Bardeen2},
so apparently the present results confirm the conjecture.
However the educated analysis of this result does not confirm the conjecture of chiral 
partnership. Notice that the model used here is known to suffer from a calibration problem. 
It is well known that in the present model the spin-orbit interaction produced by this 
potential is too large
\cite{Yaouanc}
(and that the hyperfine interaction is also too small, when compared 
with the different meson spectra). In a sense the model is too close to the starting point
of the present interpolation, the light quark chiral limit and the infinite
Isgur-Wise heavy quark limit, in the false chiral invariant vacuum, where the
spin-orbit is so large that it kills the angular splitting (and the hyperfine
and tensor potentials vanish).  If the spin-orbit interaction could be suppressed,
the splittings of eq. (\ref{chiral relation}) would increase. This increase
would easily reach the 423 MeV that separate the axialvector $D_s^*(2535)$
from the groundstate vector $D_s^*(2112)$ (if the hyperfine splitting could
be increased, the splitting between the vector $D_s^*(2112)$ and the
pseudoscalar $D_s(1968)$ would be also easily reproduced). I also notice
that, whatever these splittings turn out to be in a particular chiral
quark model, the $D$ and $D_s$ families must have similar patterns.
Then the similar 
\cite{RPP}
experimental 410 to 423 MeV mass splittings of the
vector $D^*(2007-2010)$ and axialvector $D(2420)$, 
and vector $D_s^*(2112)$ and axialvector $D_s^*(2535)$,
and the larger lattice splittings
\cite{Bali,Dougall},
all suggest that the chiral partners of the $q \bar q$ mesons 
$D_s^*(2112)$ and $D_s(1968)$ are respectively the $q \bar q$ axialvector
$D_s^*(2535)$ and a yet undetected scalar $D_s^*(2392)$. 
This educated analysis of the present results disagree with the 
beautiful and seminal conjecture  of chiral partnership for the new 
$D_s^*(2317)$ and $D_s^*(2460)$ narrow resonances. 

Importantly, this suggests that a large non $q \bar q$ component
\cite{Barnes,Terasaki}, 
say a tetraquark or a hybrid,
must be present in the new narrow $D_s$ resonances.
Coupled channels
\cite{Beveren}
or tetraquark
\cite{Bicudo_Ds}
explicit calculations,
where $D$ and $K$ mesons play a significant role,
either as a molecular state or as a coupled 
meson-meson state, also lead to the $D_s^*(2317)$ and 
$D_s^*(2460)$, and to the perfect splitting between these
mesons and the groundstates $D_s(1968)$ and $D_s^*(2112)$.

Nevertheless, once the calibration problem is solved for confining and
chiral invariant quark potentials, the techniques developed here should 
again be applied to the computation of the $D$ and $D_s$ spectra, for a 
final evaluation of the chiral partnership conjecture for the new $D_s^*$ 
mesons.

\acknowledgments

I am grateful to Maciej Nowak for motivating this
work and to George Rupp for discussions on the $D_S$
and $D$ mesons.


\end{document}